\documentclass[twocolumn]{aastex7}

\usepackage{xspace}
\usepackage{hyperref}
\usepackage[T1]{fontenc}

\newcommand{\chandra}{{\it Chandra}\xspace}
\newcommand{\xmm}{{\it XMM-Newton}\xspace}
\newcommand{\suzaku}{{\it Suzaku}\xspace}

\newcommand{\xrism}{{\it XRISM}\xspace}
\newcommand{\fexxv}{Fe\,\textsc{xxv}\xspace}
\newcommand{\fexxvi}{Fe\,\textsc{xxvi}\xspace}
\newcommand{\fexviii}{Fe\,\textsc{xviii}\xspace}
\newcommand{\fexix}{Fe\,\textsc{xix}\xspace}
\newcommand{\feka}{Fe\,K$\alpha$\xspace}
\newcommand{\fekb}{Fe\,K$\beta$\xspace}

\begin{document}

\title{A Weak Fe K{\boldmath $\beta$} Emission Line in the Broad-Line Radio Galaxy 3C~111 Observed with XRISM:\\An Ionized Wind Absorption Feature?}

\author[orcid=0000-0003-4235-5304]{Kouichi Hagino}
\affiliation{Department of Physics, The University of Tokyo, 7-3-1 Hongo, Bunkyo, Tokyo 113-0033, Japan}
\email[show]{kouichi.hagino@phys.s.u-tokyo.ac.jp}  

\author[orcid=0000-0002-2709-7338]{Motoki Kino}
\affiliation{Division of Liberal Arts, Kogakuin University of Technology \& Engineering, 2665-1 Nakano, Hachioji, Tokyo 192-0015, Japan}
\affiliation{National Astronomical Observatory of Japan, 2-21-1 Osawa, Mitaka, Tokyo 181-8588, Japan}
\email{motoki.kino@gmail.com}

\author[orcid=0000-0002-7263-7540]{{\L}ukasz Stawarz} 
\affiliation{Astronomical Observatory of the Jagiellonian University, Orla 171, 30-244 Krak\'{o}w, Poland}
\email{lukasz.1.stawarz@uj.edu.pl}

\author[orcid=0009-0003-4289-4700]{Kenzo Kawamura}
\affiliation{The University of Electro-Communications, Chofu, Tokyo 182-8585, Japan}
\affiliation{National Astronomical Observatory of Japan, 2-21-1 Osawa, Mitaka, Tokyo 181-8588, Japan}
\email{kkawamura.astro@gmail.com}

\author[orcid=0000-0001-6906-772X]{Kazuhiro Hada}
\affiliation{Graduate School of Science, Nagoya City University, Yamanohata 1, Mizuho-cho, Mizuho-ku, Nagoya 467-8501, Aichi, Japan}
\affiliation{Mizusawa VLBI Observatory, National Astronomical Observatory of Japan, 2-12 Hoshigaoka, Mizusawa, Oshu, Iwate 023-0861, Japan}
\email{hada@nsc.nagoya-cu.ac.jp}

\author[orcid=0000-0001-6020-517X]{Hirofumi Noda} 
\affiliation{Astronomical Institute, Tohoku University, 6-3 Aramakiazaaoba, Aoba-ku, Sendai, Miyagi 980-8578, Japan}
\email{hirofumi.noda@astr.tohoku.ac.jp}


\begin{abstract}
We present the results of an observation of the broad-line radio galaxy 3C~111 with the X-Ray Imaging and Spectroscopy Mission (XRISM).
The unprecedentedly high spectral resolution of XRISM/Resolve revealed that the \fekb emission line is significantly weaker than expected from the \feka line.
This feature may be explained by a blueshifted absorption line from an ionized wind overlapping the \fekb energy.
The inferred outflow velocity is 4600~km~s$^{-1}$ or 17200~km~s$^{-1}$, depending on whether the absorption feature is identified as \fexxvi or \fexxv, with the current data unable to distinguish between the two interpretations.
Based on spectral modeling, the kinetic power of the wind is estimated to lie in the range $10^{41}$--$10^{44}{\rm ~erg~s^{-1}}$, although this estimate is subject to large uncertainties primarily due to the poorly constrained location of the absorber.
The inferred wind power is smaller than the jet power of 3C~111 ($\sim 3\times 10^{44}{\rm ~erg~s^{-1}}$), and is broadly consistent with theoretical expectations that the jet power exceeds that of disk winds.
\end{abstract}



\section{Introduction}\label{sec:intro}
Radio-loud active galactic nuclei (AGNs) are a subclass of AGNs showing strong relativistic jet activity.
Approximately 10\% of AGNs are classified into this subclass depending on their radio flux compared to the optical flux, but the physical origin of the dichotomy between radio-quiet and radio-loud AGNs is not well understood~\citep[e.g.,][]{Urry1995}.
In the case of stellar mass black holes, the formation of jets is thought to be related to the black hole's spin and the magnetization of the accreting matter through the process described by \cite{Blandford1977}. 
It is also strongly affected by the state of the accretion disk and corona, as supported by observations of black hole X-ray binaries \citep[e.g.,][]{Narayan2012}.
On the other hand, in the case of AGN, the exact link between the efficiency of jet production and the states of accretion is still an open issue, leaving the physical mechanisms behind the jet dichotomy unclear.
For example, in the optical spectroscopic observations, these two classes of AGN are intrinsically similar with similar masses and accretion rates \citep{Osterbrock2006}.
Also, no clear differences were found in X-ray observations {\citep{Tazaki2013,Tombesi2014,Mestici2024}.}

The X-Ray Imaging and Spectroscopy Mission~\citep[\xrism;][]{Tashiro2025} was launched in September 2023, enabling us to investigate the environment surrounding AGNs with very high precision.
\xrism is a JAXA-led X-ray astronomical satellite developed in collaboration with NASA and ESA.
The X-ray microcalorimeter, Resolve (R. Kelley et al. in prep.), onboard \xrism has an unprecedented spectral resolution of $<5~{\rm eV}$ in full width at half maximum (FWHM) at the Fe-K band. This record-breaking resolution is a key to probing the accretion disk and surrounding materials, such as the broad line region (BLR) and torus.
Indeed, \xrism observations of the well-known Seyfert galaxy NGC 4151 revealed its nuclear structure through the detailed line shape of the neutral Fe-K$\alpha$ emission line~\citep{XRISM2024}.
Additionally, Resolve can probe the structure of disk winds launching from the very vicinity of black holes. This was clearly demonstrated by the \xrism observation of the powerful disk wind AGN PDS~456~\citep{XRISM2025}.
Thus, \xrism's great capabilities allow us to investigate the AGN structure and compare radio-loud and radio-quiet AGNs, which could help us to understand the origin of the radio-loud/quiet dichotomy.

The broad-line radio galaxy (BLRG), 3C~111 ($z=0.04884$;~\citealt{Oh2022}), is one of the most suitable targets to study the structure of the radio-loud AGNs.
It is classified as a Fanaroff-Riley class II (FR II) radio galaxy showing two radio lobes with hot spots and a single-sided jet~\citep{Linfield1984}.
From VLBI observation of superluminal motion, the viewing angle of the jet is estimated to be $\theta_{\rm view}\approx 18^\circ$ \citep[][]{Jorstad2005}.
Partly due to this moderate viewing angle, 3C 111 has been detected as a GeV $\gamma$-ray source by Fermi-LAT \citep[][]{Grandi2012}. Such detections are rare among FR II radio galaxies, since most $\gamma$-ray detected radio galaxies belong to the FR I class \citep[][]{Ackermann2015}.
While the radio and $\gamma$-ray emissions are dominated by the jet component, 3C~111 exhibits optical and X-ray emissions similar to Seyfert galaxies~\citep[e.g.,][]{Kataoka2011}.
In the optical band, a broad H$\alpha$ emission line has been detected \citep[e.g.,][]{Eracleous2003}, and its width and luminosity provide a black hole mass estimate of $1.5\textrm{--}2.4\times 10^{8} M_\odot$~\citep{Chatterjee2011}.

In the X-ray band, the Seyfert-like emission from the hot corona dominates below 10~keV, and no signatures of jet emission are observed in this energy band~\citep{Ballo2011}.
Importantly, the X-ray spectrum of 3C~111 shows a Fe-K$\alpha$ emission line at 6.4~keV \citep[e.g.,][]{Eracleous2003, Reynolds1998}, making this source the best target for studying the radio-loud AGN structure with \xrism.
In addition to the emission line, \cite{Tombesi2010b,Tombesi2011,Tombesi2013} reported the possible detections of blue-shifted Fe-K absorption lines with a velocity of $\sim0.1c$, indicating the presence of the highly ionized winds, so-called ultra-fast outflows (UFOs;~\citealt{Tombesi2010}).
Thus, \xrism observations allow us to reveal the structure of the radio-loud AGN, 3C~111, through the emission/absorption features.

In this paper, we report the results of the first \xrism observation of the BLRG 3C~111.
This paper is organized as follows.
In Section~\ref{sec:obs}, we present the \xrism observation of 3C~1111 and its data reduction procedures, and in Section~\ref{sec:result}, we describe the results of the spectral analysis of the \xrism data, including the blind search for the subtle emission/absorption line features.
In Section~\ref{sec:discussion}, we discuss our results, and finally summarize our work in Section~\ref{sec:conclusion}.

\section{Observation and Data reductions}\label{sec:obs}
3C~111 was observed by \xrism between 2024 September 25 and 2024 September 30 (OBSID: 201021010) over a total duration of 417~ks.
In this observation, Resolve was operated without a filter in the light path, and the X-ray CCD imager Xtend~\citep{Noda2025} was in 1/8 window mode to limit photon pileup.
The observed data from the Resolve and Xtend were processed and reduced using the HEAsoft package version 6.34 with the up-to-date CALDB.

The Resolve data were reduced using only high-resolution primary (Hp) events from all the pixels except for the calibration pixel (PIXEL 12) and the pixel showing large gain fluctuation (PIXEL 27).
The response matrix (RMF) and ancillary response (ARF) files were created using the tasks {\sc rslmkrmf} and {\sc xaarfgen}, respectively. 
For {\sc rslmkrmf}, we used a custom cleaned event file without the low-resolution secondary (Ls) events to prevent underestimating the RMF normalization caused by the false Ls events.
In the subsequent analysis, we used only the spectrum from 3.0 to 10.0 keV.
The source count rate was 0.845$\pm$0.002~counts~s$^{-1}$ with a net exposure of 216.3~ks.
The Resolve spectrum was binned using the optimal binning method~\cite{Kaastra2016}.
The non-X-ray background (NXB) spectrum was generated using {\sc rslnxbgen}, and modeled with a powerlaw and multiple Gaussian emission lines.
This NXB model was included for all subsequent spectral fitting, and C-statistics minimization was used.

We extracted the Xtend spectrum from a rectangular region of $5'\times 2'.26$ and the background spectrum from two rectangular regions with each size of $1'.2\times 2'.26$ in the off-source region.
The RMF and ARF files were created using the tasks {\sc xtdrmf} and {\sc xaarfgen}, respectively.
The source spectrum from 0.45 to 12.0 keV was included in the analysis.
The net background-subtracted source count rate was 4.181$\pm$0.005 counts s$^{-1}$ with a net exposure of 178.9 ks.
The Xtend spectrum was also binned using optimal binning, and the C-statistics (or exactly W-statistics) was used for the fitting.

\section{Results}\label{sec:result}
\subsection{Wide-band spectral analysis with Xtend and Resolve}
\begin{figure}[tbp]
\centering
\includegraphics[width=\hsize]{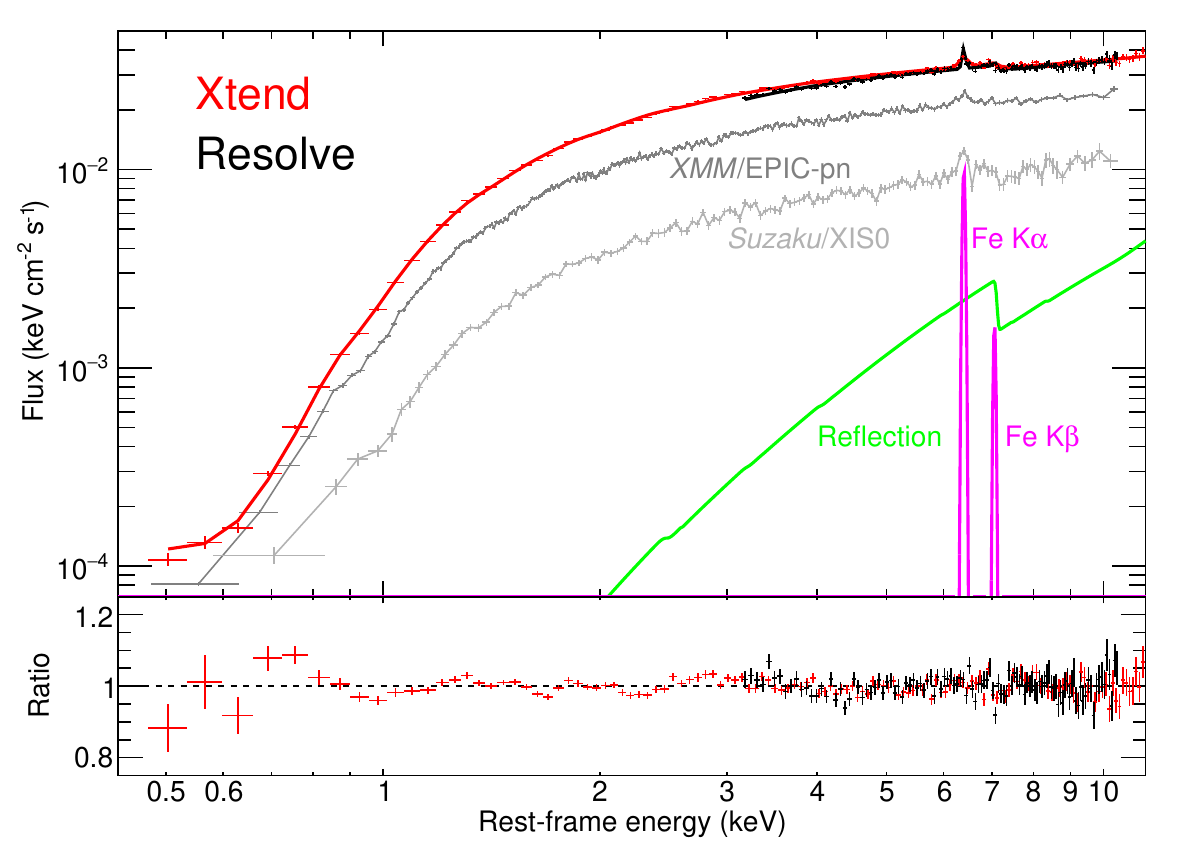}
\caption{
Wide-band \xrism spectra (Resolve in black, Xtend in red) compared with previous observations. \xmm data observed in 2009 are shown in dark grey, and \suzaku data in 2008 are in light grey. Reflection and \feka/\fekb emission line components are also shown in green and magenta, respectively.
}
\label{fig:widespec}
\end{figure}

To determine the overall continuum spectral shape of 3C~111 during this observation, we analyzed the wide-band spectra obtained by Xtend and Resolve as shown in Fig.~\ref{fig:widespec}.
As clearly seen in this figure, compared with the previous observations, the continuum curvature below 2 keV suggests a stronger absorption in this \xrism observation.
We first fitted these spectra with a simple absorbed power law and neutral Fe~K emission lines, but there was a large deviation from the data below 2 keV, with a poor fit statistic of $\textrm{C-stat/d.o.f.}=3198.12/2376$.
Then, we added an intrinsic partial-covering absorber with a column density $N_{\rm H}\sim (5\textrm{--}6)\times 10^{22}{\rm ~cm^{-2}}$ and a covering factor $f_{\rm cov}\sim 24\%$, which improved the fit statistic to 2634.36/2374.
Finally, by adding a cold reflection component with a fixed inclination angle of $\cos\theta=0.95$~\citep{Jorstad2005} as in the previous work \citep[e.g.,][]{Ballo2011,Tombesi2013}, we successfully reproduced the wide-band \xrism spectra as shown in Fig.~\ref{fig:widespec} with the best fit statistic of $\textrm{C-stat/d.o.f.}=2608.52/2373$.

\begin{deluxetable}{llc}[tbp]
\tablewidth{0pt}
\tablecaption{Best-fit parameters for the wide-band \xrism spectra \label{tab:widespec}}
\tablehead{
\colhead{Model} & \colhead{Parameter} & \colhead{Value}
}
\startdata
{\tt TBabs}   	& $N_{\rm H,gal}$ (${\rm cm^{-2}}$)	& $1.106_{-0.012}^{+0.011}\times 10^{22}$\\
{\tt zpcfabs} 	& $N_{\rm H,int}$ (${\rm cm^{-2}}$)	& $5.0_{-0.5}^{+0.5}\times 10^{22}$\\
            		& $f_{\rm cov}$			  		& $0.24_{-0.02}^{+0.02}$\\
{\tt pexrav}  	& $\Gamma$   					& $1.84_{-0.03}^{+0.03}$\\
			& $R$      					        & $0.38_{-0.12}^{+0.12}$\\
			& $N$ (${\rm ph~s^{-1}~cm^{-2}~keV^{-1}}$)   		  			& $2.35_{-0.08}^{+0.08}\times 10^{-2}$\\
{\tt zgauss} 	& $E_{\alpha1}$ (keV)						& 6.404 (fix)\\
(Fe K$\alpha_1$)			& $\sigma_{\alpha1}$ (eV)			& $28_{-8}^{+12}$\\
			& $I_{\alpha1}$ (${\rm ph~s^{-1}~cm^{-2}}$)				& $1.3_{-0.2}^{+0.2}\times 10^{-5}$\\
{\tt zgauss} 	& $E_{\alpha2}$ (keV)				& 6.391 (fix)\\
(Fe K$\alpha_2$)		 	& $\sigma_2$ (eV)			& $=\sigma_{\alpha1}$\\
			& $I_{\alpha2}$ (${\rm ph~s^{-1}~cm^{-2}}$) 				& $=I_{\alpha1}/2$\\
{\tt zgauss} 	& $E_{\beta}$ (keV)				& 7.058 (fix)\\
(Fe K$\beta$)		 	& $\sigma_{\beta}$ (eV)			& $=\sigma_{\alpha1}$\\
			& $I_{\beta}$ (${\rm ph~s^{-1}~cm^{-2}}$) 				& $=0.135\times (I_{\alpha1}+I_{\alpha2})$\\
\hline
  	& C-stat/d.o.f. 					& 2608.52/2373 \\
\enddata
\end{deluxetable}

\begin{figure*}[tbp]
\centering
\includegraphics[width=0.49\hsize]{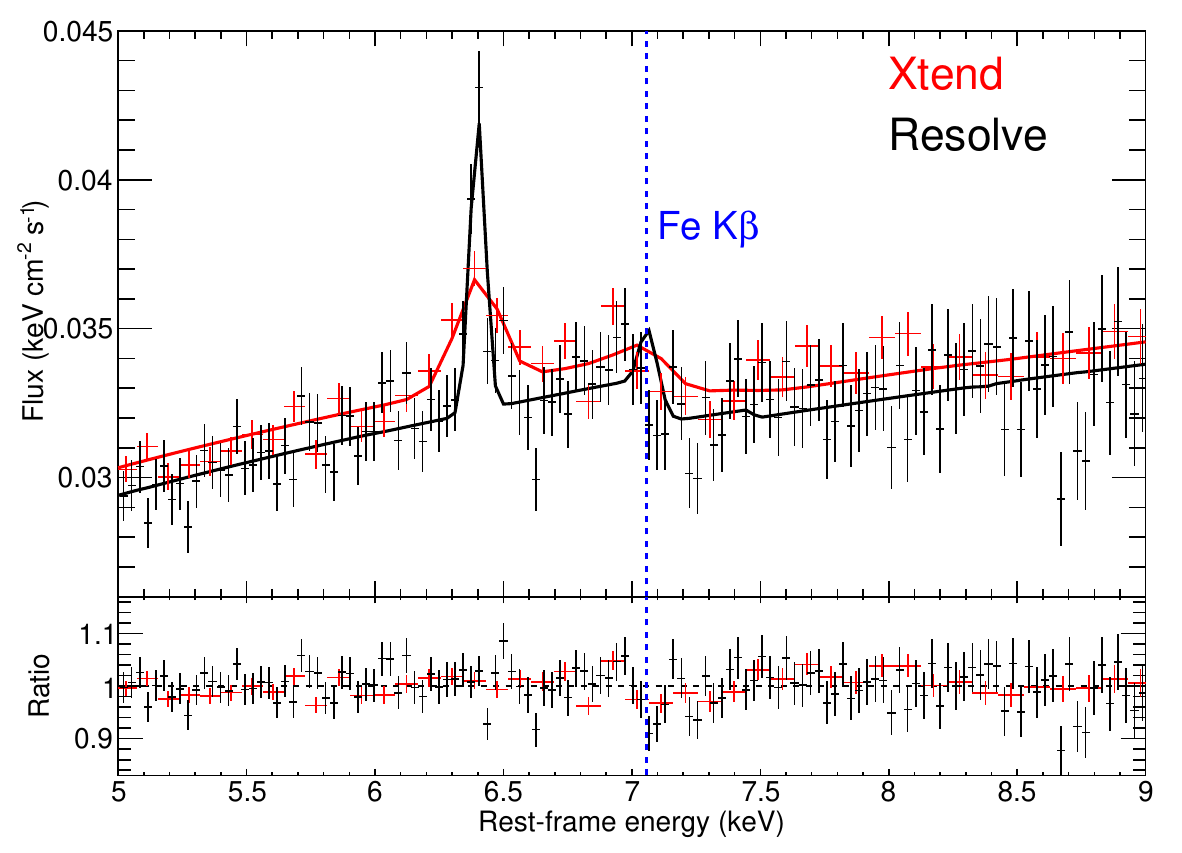}
\includegraphics[width=0.49\hsize]{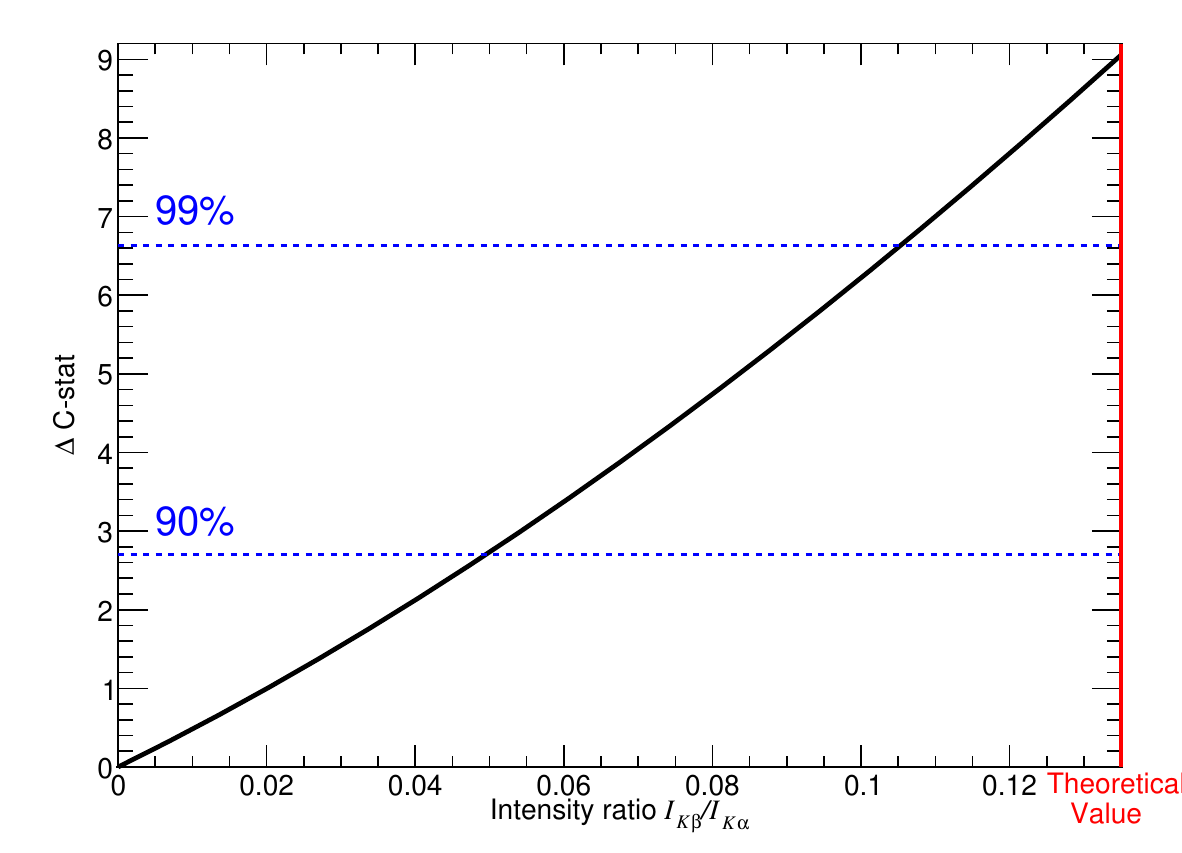}
\caption{
({\it Left}) 5--9 keV \xrism spectra of 3C~111. A vertical blue dashed line indicates the \fekb energy, where a clear residual exists.
({\it Right}) One-dimensional contour plot of $\Delta$C-stat against the intensity ratio between \fekb and \feka.
}
\label{fig:kb}
\end{figure*}

Table~\ref{tab:widespec} summarizes the best-fit parameters for the 0.45--12.0~keV Xtend and 3.0--10.0~keV Resolve spectra.
In {\sc xspec} terminology, the model is described as {\tt TBabs*zpcfabs(gsmooth*pexrav + zgauss + zgauss + zgauss)}, where {\tt TBabs} is a Galactic absorption and {\tt zpcfabs} is an intrinsic partial covering absorber located at the source rest frame of $z=0.04884$.
The {\tt gsmooth*pexrav} component is a cut-off powerlaw continuum with a cold reflection, convolved with an additional Gaussian smoothing with the same width as emission lines. The three {\tt zgauss} components are Gaussians that model the \feka ﬂuorescence emission doublet (K$\alpha_1$ at 6.404~keV and K$\alpha_2$ at 6.391~keV) and \fekb line at 7.058~keV from neutral material.
Additionally, we also multiplied a free cross-normalization constant of 2\% between Xtend and Resolve.

In this fitting, both a Galactic absorption with $N_{\rm H,gal}=1.106\times 10^{22}{\rm ~cm^{-2}}$ and an intrinsic absorption with $N_{\rm H,int}=5.0\times 10^{22}{\rm ~cm^{-2}}$ are required.
This Galactic column density clearly exceeds the estimates from 21~cm radio surveys of atomic H\,\textsc{i}, but it is fully consistent with the previous X-ray observations~\citep[e.g.,][]{Lewis2005,Ballo2011}.
This excess is probably due to the molecular gas associated with the Taurus molecular cloud complex toward 3C~111 \citep{Tombesi2017}.
The additional intrinsic absorber had a larger column density { of $N_{\rm H,int}=5.0\times 10^{22}{\rm ~cm^{-2}}$} with a partial covering factor $f_{\rm cov}=24\%$.
{ 
Similar intrinsic neutral absorbers were reported in the 2010 \suzaku data, but the column density was much smaller at $6.5\times 10^{21}{\rm ~cm^{-2}}$ \citep{Tombesi2013}.
Moreover, no such intrinsic absorptions were required in the 2008 \suzaku or the 2009 \xmm data \citep{Ballo2011}.
Thus, as can be seen from the difference in the low-energy curvature in Fig.~\ref{fig:widespec}, 3C~111 was in an exceptionally high intrinsic absorption phase during our \xrism observation.
We note that this partial-covering neutral absorber model may be degenerate with a scenario including a fully covering neutral absorber plus a scattered continuum component.}

The intrinsic power-law component was found to have a relatively hard photon index of $\Gamma=1.84$ compared with previous observations.
This is consistent with the correlation between the photon index $\Gamma$ and 2--10 keV flux $F_{\rm 2\textrm{--}10~keV}$ reported in \cite{Chatterjee2011} because of the larger unabsorbed flux of $F_{\rm 2\textrm{--}10~keV}=7.9\times 10^{-11}{\rm ~erg~cm^{-2}~s^{-1}}$ in our \xrism observation.
The cold reflection component was also required to explain the observed continuum spectrum with $\Delta\textrm{C-stat}=25.84$ for only one additional parameter.
The necessity of the reflection component was also clearly seen in its edge feature shown in the left panel of Fig.~\ref{fig:kb}.
The best-fit reflection fraction $R=0.38$ is larger than most of the previous observations, but consistent with 2008 Suzaku data~\citep{Ballo2011}.

The narrow \feka emission line was clearly detected in both Xtend and Resolve spectra.
Thanks to the unprecedentedly high spectral resolution of Resolve, we successfully measured the very narrow width of $\sigma=28{\rm ~eV}$, which was not resolved in the previous instruments such as CCDs and grating instruments~\citep{Tombesi2017}.
The total intensity of Fe~K$\alpha_1$ and K$\alpha_2$ was $2.0\times 10^{-5}{\rm ~ph~s^{-1}~cm^{-2}}$.
It was consistent with the previous observations with \xmm, \suzaku, and \chandra/HETG \citep{Ballo2011,Tombesi2017}

As shown in the left panel of Fig.~\ref{fig:kb}, a comparison of Resolve and Xtend spectra with the best-fit model shows a clear residual feature at 7.0--7.1~keV, where the \fekb emission line is located.
Since this feature coincides with the \fekb energy, we treated the \fekb intensity as a free parameter and tested the statistical improvement of the fit.
The right panel of Fig.~\ref{fig:kb} shows the improvement of the fit statistics $\Delta$C-stat with the \fekb intensity.
The best-fit value of the \fekb intensity is zero, i.e., no \fekb emission line is favored.
Based on this fit, the theoretically-expected value $I_{\rm K\beta}/I_{\rm K\alpha}=0.135$ \citep{Palmeri2003,Yaqoob2010} is rejected more than 99\% of confidence level.

\subsection{Wind modeling for the Resolve Fe~K spectrum}
\begin{figure}[tbp]
\centering
\includegraphics[width=\hsize]{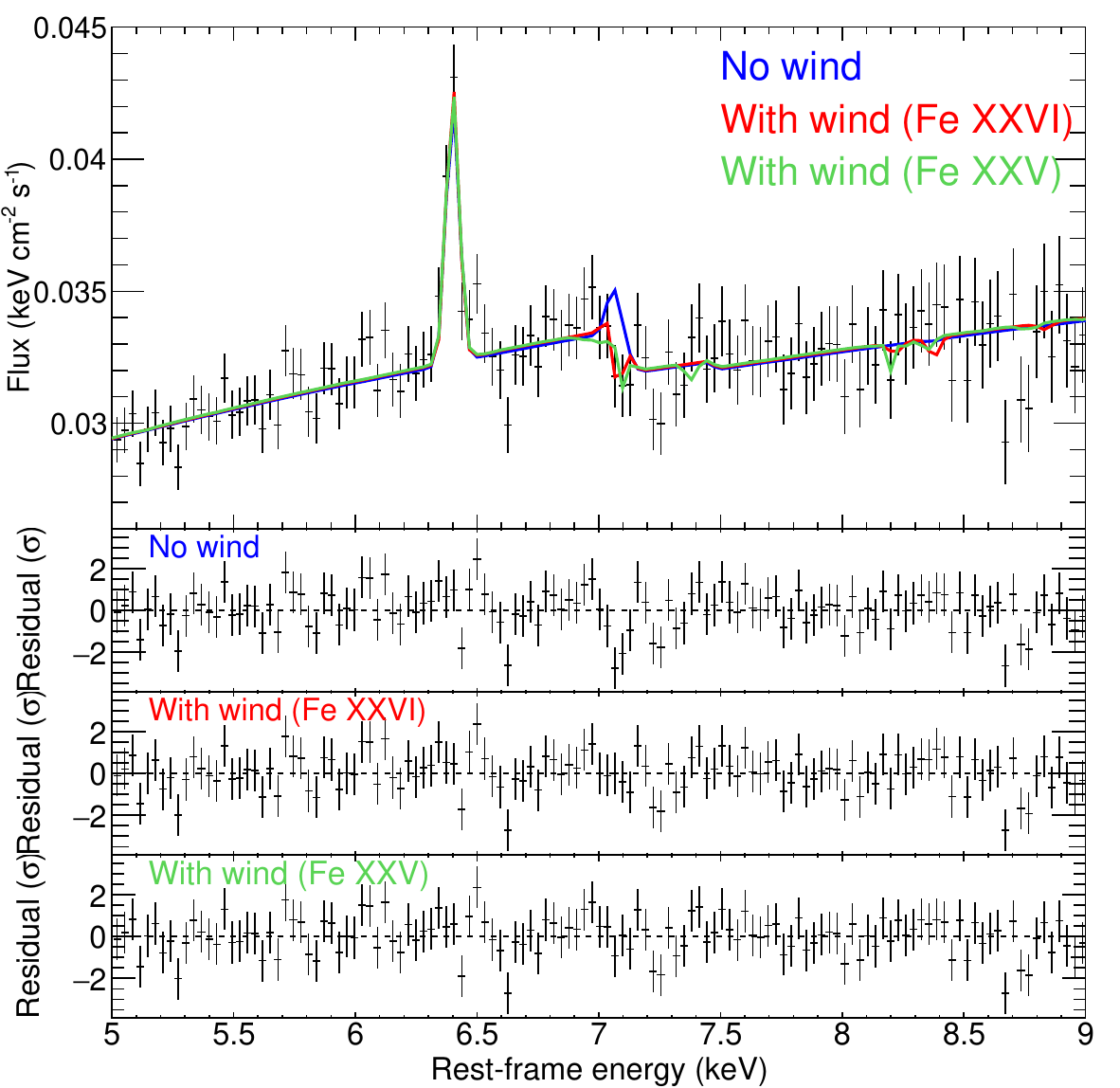}
\caption{
Spectral fitting of the Resolve spectrum at 5--9~keV using spectral models with wind components.
These models mainly attribute the absorption line to \fexxvi or \fexxv.
}
\label{fig:fek_wind}
\end{figure}

The most feasible explanation for the unexpectedly weak \fekb emission line is a wind absorption line overlapping at the \fekb emission energy.
To test this hypothesis, we first phenomenologically fit the data by adding a negative Gaussian component.
The residual at the \fekb energy is successfully reduced with this additional Gaussian with a width of $27^{+14}_{-10}$~eV centered at $7.074^{+0.018}_{-0.015}$~keV.
The best-fit equivalent width (EW) is $-8^{+3}_{-4}$~eV, which is much weaker than the previously detected wind feature with an EW of $-26\pm6$~eV~\citep{Tombesi2011}.
The fit statistic is improved by $\Delta\textrm{C-stat}/\Delta{\rm d.o.f}=15.28/3$, corresponding to a p-value of 0.0016.
It indicates that this absorption feature is significant with a more than $3\sigma$ confidence level.

We used a more physical model for the wind absorption component.
In this analysis, we fitted only the Resolve spectrum, while fixing the continuum parameters at the best-fit value listed in Table~\ref{tab:widespec} except for the normalization and reflection fraction.
This analysis allows us to avoid the systematic biases caused by Xtend's low-resolution, high-statistic data during the fitting process.
The wind absorption component was modeled with the XSTAR photoionization code~\citep{Kallman2004}.
The ionizing spectrum was assumed as a simple power-law with a photon index of 1.8 based on the best-fit continuum slope.
In the fitting, the ionization parameter $\xi$, outflow velocity $v$, column density $N_{\rm H}$, and the velocity width $\sigma_{\rm v}$ were treated as free parameters~\citep{Mochizuki2023}.

As shown in Fig.~\ref{fig:fek_wind}, we found two statistically indistinguishable solutions.
In one solution, the absorption line is primarily attributed to \fexxvi, while in the other, it is attributed to \fexxv.
The best-fit parameters are listed in Table~\ref{tab:fekspec}.
In the \fexxvi model, ionization parameter and column density were obtained as lower limits of $\log\xi>3$ and $N_{\rm H}> 0.5\times 10^{22}{\rm ~cm^{-2}}$.
On the other hand, in the \fexxv model, these parameters are well determined to be $\log\xi\simeq 2.8$ and $N_{\rm H}\simeq 0.5\times 10^{22}{\rm ~cm^{-2}}$.
This is because only the \fexxv line must be stronger than the other lines to explain the observed feature.
The outflow velocity $v$ was determined as $v\simeq 4600{\rm ~km~s^{-1}}$ and $\simeq17200{\rm ~km~s^{-1}}$ for \fexxvi and \fexxv models, respectively.

\begin{deluxetable}{lcc}[tbp]
\tablewidth{0pt}
\tablecaption{Best-fit wind parameters for the Resolve spectrum \label{tab:fekspec}}
\tablehead{
\colhead{} & \colhead{\fexxvi model} & \colhead{\fexxv model}
}
\startdata
$\log\xi$ (erg~cm~s$^{-1}$) & $>3.0$ & $2.8_{-0.2}^{+0.2}$ \\
$N_{\rm H}$ (cm$^{-2}$) & $>0.5\times 10^{22}$ & $0.5_{-0.2}^{+0.3} \times 10^{22}$ \\
$v$ (km~s$^{-1}$) & $-4600_{-500}^{+700}$ & $-17200_{-400}^{+1300}$\\
& ($-0.0153_{-0.0017}^{+0.0023}c$) & ($-0.0573_{-0.0013}^{+0.0043}c$) \\
$\sigma_{\rm v}$ (km~s$^{-1}$) & $<2000$ & $<1300$ \\
\hline
C-stat/d.o.f. & 2322.26/2231 & 2322.28/2231 \\
$\Delta$C-stat/$\Delta$d.o.f. & 12.38/4 & 12.36/4 \\
p-value & 0.015 & 0.015\\
\enddata
\end{deluxetable}

{
Additionally, we performed a blind line search to check the significance of the 7.07~keV absorption line and to identify other possible line features.
Similarly to the above analysis, we fitted only the Resolve spectrum with fixed continuum parameters, except for the normalization and reflection fraction.
The neutral Fe~K emission lines were also included with a fixed intensity ratio of 0.135.
To search for line features, an additional Gaussian line with fixed line energy and width (20~eV and 50~eV) was added, and the statistical improvement was recorded.
Both positive and negative line normalizations were allowed.
The line energy was scanned over 6.6--10.5~keV in the rest frame with a step size of 1~eV.
As a result, we evaluated the local significance of additional line features at rest-frame energies from 6.6~keV to 10.5~keV, as shown in Fig.~\ref{fig:linesearch}.
This result confirmed that the 7.07~keV absorption line is significant with $3\sigma$ confidence level without considering the look-elsewhere effect.
In addition, three marginal features are found at the $2\sigma$ significance level at 6.96, 7.24, and 9.66~keV.
The 6.96~keV emission feature is possibly related to the He-like Fe~K emission line, and 7.24 and 9.66~keV absorption features may be associated with additional absorption components from the ionized outflow, although the current statistical significance is limited.

\begin{figure}[tbp]
\centering
\includegraphics[width=\hsize]{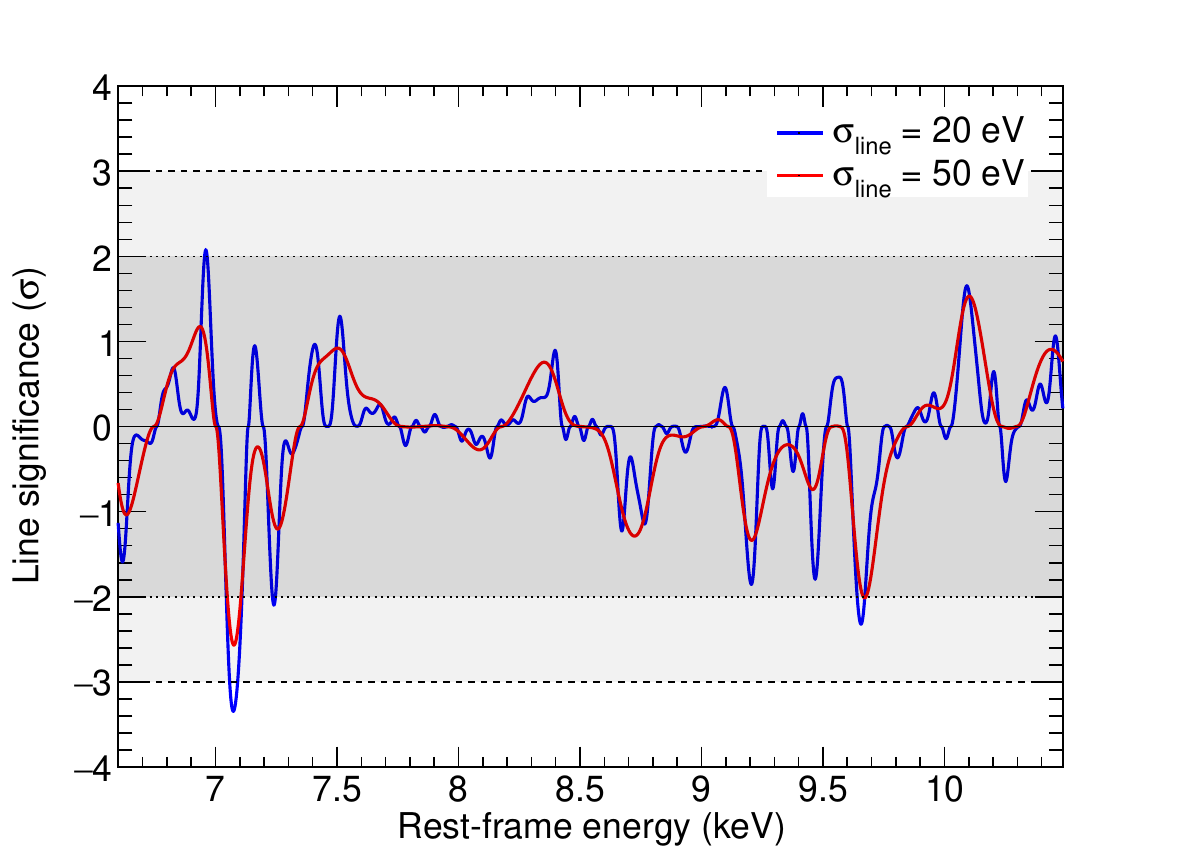}
\caption{
Line significance in units of $\sigma$, obtained from a blind line search in the 6.6--10.5 keV Resolve spectrum with a Gaussian line width of $\sigma_{\rm line} = 20$~eV (blue) and 50~eV (red).
Positive and negative significance values indicate emission and absorption lines, respectively.
In the 20~eV search, the absorption feature at 7.07~keV exceeds $3\sigma$ confidence level.
}
\label{fig:linesearch}
\end{figure}
}

\section{Discussions}\label{sec:discussion}
\subsection{Alternative interpretation of the weak \fekb line}
\begin{figure}[tbp]
\centering
\includegraphics[width=\hsize]{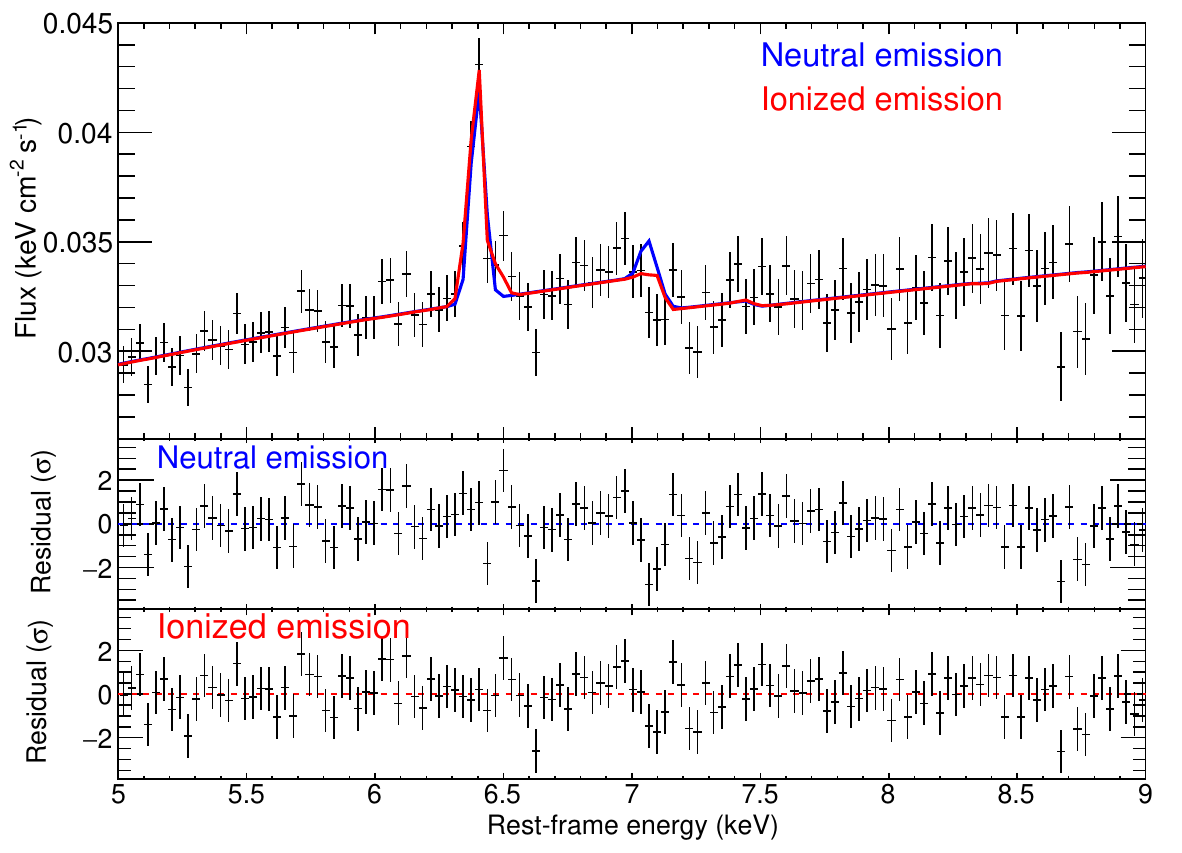}
\caption{
Spectral fitting of the Resolve spectrum at 5--9~keV using a spectral model with ionized emission lines instead of the neutral emission lines.
}
\label{fig:fek_ion}
\end{figure}

Not only wind absorption, but also emission lines in a moderate ionization state can explain weak \fekb lines.
Since the K$\beta$ line is emitted due to the transition from M-shell to L-shell, if ions have few M-shell electrons, the intensity of the K$\beta$ line is suppressed~\citep[e.g.,][]{Yamaguchi2014}.
To quantitatively test this scenario, we modeled the ionized emission lines with XSTAR and fitted the Resolve spectrum with a model composed of ionized emissions instead of neutral emission lines.
Fig.~\ref{fig:fek_ion} shows the best-fit model with the ionized emission lines.
As expected, the intensity of the \fekb line was clearly suppressed with an ionization parameter of $\log\xi\simeq2.0$, where \fexviii and \fexix are dominant.
However, to explain the observed data, an additional redshift with a velocity of $3000~{\rm km~s^{-1}}$ is required because \feka lines of \fexviii and \fexix are located at 30--70~eV higher energy than neutral \feka emission line.
This indicates that the emitting material is moving away from the observer, suggesting an inflow.
Also, iron abundance needs to be more than 1.4 times the solar abundance to suppress the Ca K$\alpha$ emission line that is not seen in the observed data.
This ionized emission line needs to have a much narrower line width of 10~eV than the neutral emission line width of 28~eV because the \feka line is composed of two lines with a separation of 40~eV in this scenario.

This moderately ionized emission scenario is unlikely in terms of the geometrical structure and the conventional view of AGN structure.
This ionized emission is redshifted with a velocity of $3000{\rm ~km~s^{-1}}$, but the line width was only $\sigma\sim 500{\rm ~km~s^{-1}}$.
Since some fraction of this width should be caused by the {Keplerian} motion of the emitting material, the emitting material is located at a radius of $\gtrsim 10^5R_{\rm g}$ from the black hole.
In addition to the Keplerian motion, the dispersion of the inflow direction should also contribute to the emission width.
If the {emitting} material inflows with an angle of $\theta$ against the disk rotation axis, the line-of-sight velocity $v$ at an inclination angle of $i=18^\circ$ and azimuthal angle $\phi$ is written as $v(\phi)=v_0(\sin\theta\sin i\cos\phi + \cos\theta\cos i)$.
Averaging over $\phi=0\textrm{--}2\pi$, the mean line-of-sight velocity is $\bar{v}=v_0\cos\theta\cos i$ and the standard deviation is $\sigma_v=v_0\sin\theta\sin i/\sqrt{2}$.
Considering $500{\rm ~km~s^{-1}}$ as an upper limit of $\sigma_v$, $\sigma_v/\bar{v}=\tan\theta\tan i/\sqrt{2}<500{\rm ~km~s^{-1}}/3000{\rm ~km~s^{-1}} \Leftrightarrow \theta \lesssim 36^\circ$ is obtained.
Therefore, the moderately ionized material has a peculiar geometrical structure that inflows at vertical direction with $\theta \lesssim 36^\circ$ at a relatively large radius of $\gtrsim 10^5R_{\rm g}$.
Also, if the Fe~K emission line is dominated by this moderately ionized emission instead of BLR or torus, the nuclear structure of 3C~111 is different from the normal Seyfert galaxies, contradicting to previous work that found no clear differences between radio galaxies and Seyfert galaxies in X-ray observations~\citep{Tazaki2013}.
Consequently, the ionized wind scenario is a much more realistic scenario than the moderately ionized emission scenario.

\subsection{Wind location and energetics}
Based on the results of wind modeling, we estimate the wind location.
The maximum radial distance from the X-ray source can be estimated as $R_{\rm max}=L_{\rm ion}/(N_{\rm H}\xi)$ by assuming $\Delta R<R\Leftrightarrow N_{\rm H}<nR$. Here, the ionizing luminosity in 1--1000~Ry ($=0.0136\textrm{--}13.6{\rm ~keV}$) is estimated to be $L_{\rm ion}\simeq1.3\times 10^{45}{\rm ~erg~s^{-1}}$ by extrapolating the observed X-ray continuum.
On the other hand, the minimum radial distance can be estimated as $R_{\rm min}=2/(v/c)^2R_{\rm g}$ if the wind can escape from the gravity, i.e., not a failed wind.
The gravitational radius $R_g=GM/c^2$ is estimated as $\simeq 3\times 10^{13}{\rm ~cm}$ for the black hole mass of $M\simeq 2\times 10^8{\rm ~M_\odot}$.
Using the best-fit parameters, the radial distances of the wind are estimated to be 
$R\simeq 0.07\textrm{--}100{\rm ~pc}\simeq 7\times 10^3 \textrm{--}9\times 10^6R_{\rm g}$
for \fexxvi model
and
$R\simeq 0.007\textrm{--}300{\rm ~pc} \simeq 6\times 10^2 \textrm{--}4\times 10^7R_{\rm g}$
for \fexxv model.
The upper limits of the location are very large as $\sim10^7R_{\rm g}\sim 100{\rm~pc}$ due to the observed small column density of the wind.

We also estimate the mass outflow rate $\dot{M}_{\rm out}$ based on the estimated location.
It is given as $\dot{M}_{\rm out}=\Omega \mu m_{\rm p}nR^2v$, where $\Omega$ is the solid angle of the wind, $\mu\simeq 1.4$ is the mean atomic mass per proton, and $m_{\rm p}$ is the proton mass.
Assuming that the wind distributes from $R_{\rm in}$ to infinity with a constant mass outflow rate $\dot{M}_{\rm out}$ and outflow velocity $v$, the mass outflow rate can be estimated as $\dot{M}_{\rm out}=\Omega \mu m_{\rm p}N_{\rm H}Rv$~\citep[e.g.,][]{Nardini2015}.
As a result, the mass outflow rates are calculated to be 0.009--10$\rm ~M_\odot~yr^{-1}$ for \fexxvi model and 0.003--160$\rm ~M_\odot~yr^{-1}$ for \fexxv model, where the solid angle of the wind is assumed as $\Omega/4\pi=0.5$~{\citep{Tombesi2010, Gofford2013}.}
However, these upper bounds of mass outflow rate are physically unlikely, considering the balance with mass inflow rate $\dot{M}_{\rm in}$.
Using the relation of $\dot{M}_{\rm in}=L_{\rm bol}/\eta c^{2}$, where $L_{\rm bol}$ and $\eta$ are the bolometric luminosity and the radiative efficiency, respectively, we estimate  $\dot{M}_{\rm in}$ as follows:
\begin{eqnarray}
\dot{M}_{\rm in}= 0.176~
\left(\frac{L_{\rm bol}}{10^{45}~{\rm erg~s^{-1}}}\right)
\left(\frac{\eta}{0.1}\right)^{-1}
{\rm ~M_{\odot}~yr^{-1}},
\end{eqnarray}
together with the uncertainty of $0.06 \leq \eta_{\rm rad} \leq 0.15$ both theoretically and observationally \citep[][]{Davis2011, McKinney2015}.
Taking into account the uncertainty in $L_{\rm bol}$, $2.5 \times 10^{44}~{\rm erg~s^{-1}} \leq L_{\rm bol}\leq 8.8\times 10^{45}~{\rm erg~s^{-1}}$ \citep[][]{Marchesini2004,Ballo2011, DeJong2012, Tombesi2013}, we obtain the mass accretion rate as
$0.03~{\rm M_{\odot}~yr^{-1}}
\lesssim \dot{M}_{\rm in}\lesssim 
2.6~{\rm M_{\odot}~yr^{-1}}$.
Since the mass outflow rate does not exceed the mass inflow rate in general, we can set the upper limit of the mass outflow rate as $\dot{M}_{\rm out}\lesssim 2.6~{\rm M_{\odot}~yr^{-1}}$.
Thus, the estimated mass outflow rate ranges as:
\begin{eqnarray*}
0.009~{\rm M_{\odot}~yr^{-1}} \lesssim &\dot{M}_{\rm out}&\lesssim 2.6~{\rm M_{\odot}~yr^{-1}}{\rm ~(\fexxvi)}\\
0.003~{\rm M_{\odot}~yr^{-1}} \lesssim &\dot{M}_{\rm out}&\lesssim 2.6~{\rm M_{\odot}~yr^{-1}}{\rm ~(\fexxv)}.
\end{eqnarray*}
This upper limit inversely sets the upper limits of the radial location, resulting in 
\begin{eqnarray*}
7\times 10^3R_{\rm g} \lesssim &R&\lesssim 2\times 10^6R_{\rm g}{\rm ~(\fexxvi)}\\
6\times 10^2R_{\rm g} \lesssim &R&\lesssim 7\times 10^5R_{\rm g}{\rm ~(\fexxv)}.
\end{eqnarray*}

Finally, we estimate the wind kinetic power $P_{\rm wind}=\dot{M}_{\rm out}v^2/2$.
It is estimated to be 
\begin{eqnarray*}
6\times10^{40}{\rm ~erg~s^{-1}} \lesssim &P_{\rm wind}&\lesssim 2\times10^{43}{\rm ~erg~s^{-1}}{\rm ~(\fexxvi)}\\
3\times10^{41}{\rm ~erg~s^{-1}} \lesssim &P_{\rm wind}&\lesssim 2\times10^{44}{\rm ~erg~s^{-1}}{\rm ~(\fexxv)}.
\end{eqnarray*}
This estimated wind kinetic power is smaller than the jet power of 3C~111 $P_{\rm jet}\simeq 3\times 10^{44}{\rm ~erg~s^{-1}}$ based on the assumption of equipartition~\citep{Tombesi2012a}.
This agrees well with the numerical simulation results that suggest a relation of $P_{\rm jet}\gtrsim 10P_{\rm wind}$~\citep{Sadowski2013, Yang2021}.

{
The Eddington ratio of 3C~111 is estimated to be $L_{\rm bol}/L_{\rm Edd}\simeq 0.01\textrm{--}0.35$ by using the Eddington luminosity $L_{\rm Edd}\simeq2.5\times10^{46}{\rm ~erg~s^{-1}}$ for the black hole mass of $M\simeq 2\times 10^8{\rm ~M_\odot}$.
Thus, 3C~111 is accreting in the sub-Eddington regime.
Normalized by the Eddington luminosity, the estimated wind kinetic power corresponds to $P_{\rm wind}/L_{\rm Edd}\simeq 2\times10^{-6}\textrm{--}8\times10^{-3}$, indicating that the wind is energetically modest compared with the total accretion power.
In contrast, the jet power reaches $P_{\rm jet}/L_{\rm Edd}\sim10^{-2}$, suggesting that the jet is likely the dominant channel of kinetic energy output in 3C~111.
}

\subsection{Connection with jet activity}
\begin{figure}[tbp]
\centering
\includegraphics[width=\hsize]{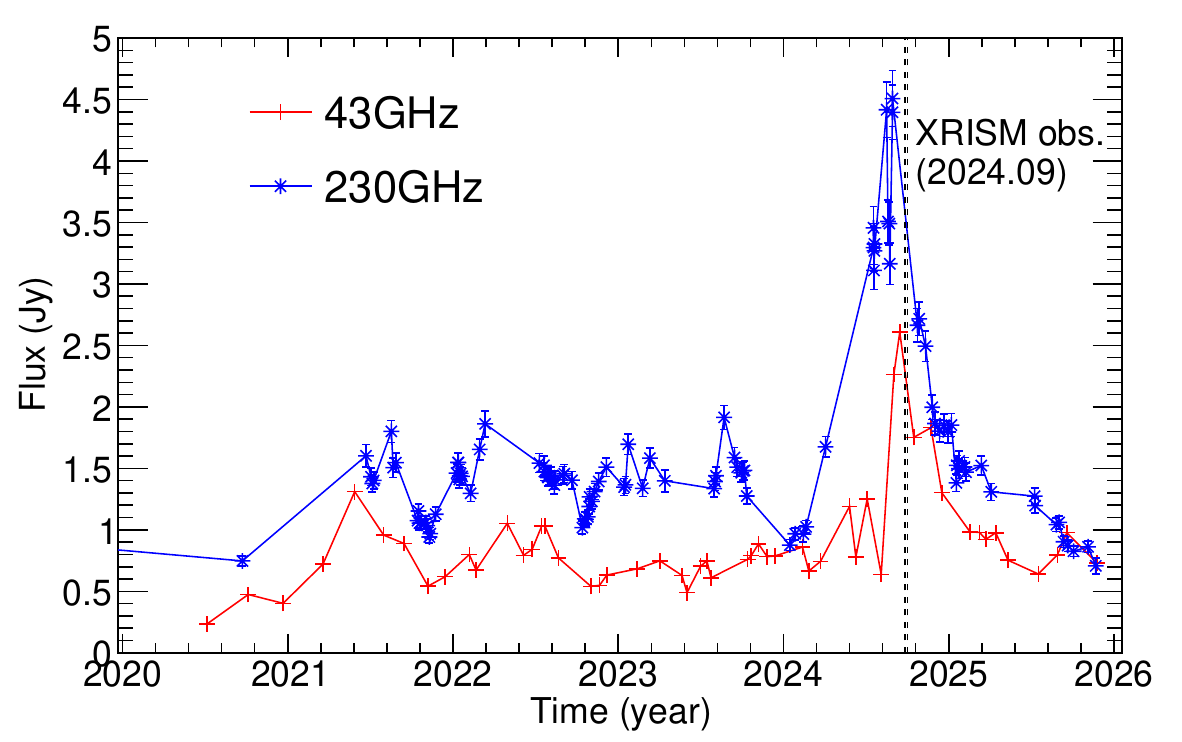}
\caption{
Long-term radio light curves at 43~GHz and 230~GHz of 3C~111 from 2020 to 2026.
The vertical dashed line indicates the date of the \xrism observation.
}
\label{fig:radio_lc}
\end{figure}

X-ray activity, often regarded as a tracer of accretion disk activity, is known to be connected to jet activity in radio-loud AGN~{\citep{Marscher2002,Chatterjee2009,Chatterjee2011,Tombesi2014,Mestici2024}.}
Disk winds have also been detected in radio-loud AGN, suggesting that they can coexist with relativistic jets \citep{Tombesi2010,Gofford2013}.
The interplay between disk winds and jets has been discussed in the framework of magnetically driven outflows \citep{Fukumura2010,Fukumura2014}.
In 3C~111, \cite{Tombesi2012a} suggested that disk winds may also be related to this disk-jet connection.
In their scenario, dips in the X-ray flux are interpreted as temporary disappearances of the inner accretion disk, which are followed by the ejection of bright superluminal knots in the radio jet on timescales of a few months.
They reported that winds were detected only during intervals of increasing X-ray flux after the dip, suggesting that strong instabilities associated with the knot ejection may trigger or enhance the winds.

To examine whether the possible wind detected in our \xrism observation is consistent with this disk-jet-wind connection, we investigated the timing of the \xrism observation relative to the radio jet activity.
As a proxy for jet activity, we used publicly available radio monitoring data from the VLBA at 43~GHz provided by Boston University\footnote{\url{https://www.bu.edu/blazars/BEAM-ME.html}} and from the Submillimeter Array at Mauna Kea (230~GHz).
The  230 GHz flux density data for 3C111 was obtained at the Submillimeter Array (SMA) near the summit of Maunakea (Hawaii). 3C111 is included in an ongoing monitoring program at the SMA to determine the ﬂuxes of compact extragalactic radio sources that can be used as calibrators at mm wavelengths \citep{Gurwell2007}.  Observations of available potential calibrators are from time to time observed for 3 to 5 minutes, with the measured source signal strength calibrated against known standards, typically solar system objects (Titan, Uranus, Neptune, or Callisto).  Data from this program are updated regularly and are available at the SMA website\footnote{\url{http://sma1.sma.hawaii.edu/callist/callist.html}}.
Fig.~\ref{fig:radio_lc} shows the long-term radio light curves of 3C~111 from 2020 to 2026.
As shown in this figure, the \xrism observation was performed during a phase when the radio flux was gradually decreasing after a flare.
According to the scenario proposed by \cite{Tombesi2012a}, strong winds are preferentially detected during the flux-increasing phase following an X-ray dip, whereas no clear wind signatures were reported during the decreasing phase.

The \suzaku observation on 2010 September 14 (Obs~3 in \citealt{Tombesi2011}) provides a useful comparison because it was also carried out during a similar decreasing-flux interval.
That observation did not show any significant wind feature and provided only a lower limit on the equivalent width of the absorption line (${\rm EW} > -19$~eV).
In contrast, our \xrism observation reveals a weak absorption feature with an equivalent width of $-8$~eV, which is consistent with the sensitivity limit of the earlier observation.
Although the exact timing of the X-ray dip cannot be determined with current data due to a lack of long-term X-ray monitoring below 10~keV like RXTE, the weak wind feature detected by \xrism does not contradict the disk-jet-wind connection proposed by \cite{Tombesi2012a}.
Future simultaneous X-ray and VLBI observations will be crucial to clarify the temporal relationship between disk winds and jet ejection events in this source.

\section{Conclusions}\label{sec:conclusion}
We performed {a} \xrism observation of the broad-line radio galaxy 3C~111 with net exposures of 216.3~ks for Resolve and 178.9~ks for Xtend.
High-resolution spectroscopy with Resolve revealed that the \fekb emission line is significantly weaker than expected from the \feka line intensity.
One possible explanation for this feature is a blueshifted absorption line from highly ionized gas, identified as either \fexxvi or \fexxv, overlapping the \fekb energy.
Spectral modeling with XSTAR yielded two statistically indistinguishable solutions corresponding to these two identifications.
The implied wind velocity is 4600~km~s$^{-1}$ for the \fexxvi interpretation and 17200~km~s$^{-1}$ for the \fexxv interpretation, and the current data do not allow us to discriminate between them.
The absorber is inferred to be highly ionized, with an ionization parameter of $\log\xi\gtrsim 3$, and to have a relatively small column density of $N_{\rm H}\gtrsim 0.5\times10^{22}{\rm cm^{-2}}$.
The location of the absorbing gas is weakly constrained, spanning $\sim 7\times10^3\textrm{--}2\times10^6R_{\rm g}$ for the \fexxvi model and $\sim 6\times10^2\textrm{--}7\times10^5R_{\rm g}$ for the \fexxv model, which leads to a large uncertainty in the derived kinetic power of the wind, estimated to be $10^{41}$--$10^{44}{\rm ~erg~s^{-1}}$.
This wind power is lower than the jet power of 3C~111 ($\sim 3\times 10^{44}{\rm ~erg~s^{-1}}$), and is broadly consistent with theoretical expectations that the jet power exceeds that of disk winds.

\begin{acknowledgments}
We thank M. Gurwell for kindly providing the SMA data.
This work was supported by JSPS KAKENHI grant numbers JP21K13963, JP24K00638, JP26K12345, JP25H00660, JP22H00157, and JP21H04488.
K.K. acknowledges support from JST SPRING, Grant Number JPMJSP2131.
K.Hada is also supported by Mitsubishi Foundation (grant 202310034) and Daiko Foundation (grant J0SE807004). This work was supported in part by a University Research Support Grant from the National Astronomical Observatory of Japan (NAOJ), and by Grant-in-Aid for Outstanding Research Group Support Program in Nagoya City University Grant Number 2530002.
The Submillimeter Array is a joint project between the Smithsonian Astrophysical Observatory and the Academia Sinica Institute of Astronomy and Astrophysics and is funded by the Smithsonian Institution and the Academia Sinica.
We recognize that Maunakea is a culturally important site for the indigenous Hawaiian people; we are privileged to study the cosmos from its summit.
This study makes use of VLBA data from the VLBA-BU Blazar Monitoring Program (BEAM-ME and VLBA-BU-BLAZAR;
\url{http://www.bu.edu/blazars/BEAM-ME.html}), funded by NASA through the Fermi Guest Investigator Program. The VLBA is an instrument of the National Radio Astronomy Observatory. The National Radio Astronomy Observatory is a facility of the National Science Foundation operated by Associated Universities, Inc.
\end{acknowledgments}





%
\facilities{XRISM, SMA}



%

\bibliography{ref}{}
\bibliographystyle{aasjournalv7}



\end{document}